\documentclass[12pt]{article}
\usepackage{amsmath,amsfonts}
\usepackage[english]{babel}
\usepackage{epic,latexsym}
\usepackage[dvips]{graphicx}
\usepackage[latin1]{inputenc}
\usepackage[usenames]{color}
\usepackage{setspace}
\usepackage{natbib,vmargin}

\catcode`@=11 
\def\section{\@startsection {section}{1}{\z@}{-3.5ex plus -1ex minus
 -.2ex}{2.3ex plus .2ex}{\normalsize\bf\centering}}
\def\subsection{\@startsection{subsection}{2}{\z@}{-3.25ex plus -1ex minus
 -.2ex}{1.5ex plus .2ex}{\normalsize\bf}}
\def\subsubsection{\@startsection{subsubsection}{3}{\z@}{-3.25ex plus
 -1ex minus -.2ex}{1.5ex plus .2ex}{\normalsize\it}}
\catcode`@=12 

\setpapersize{USletter}
\setmarginsrb{.9in}{1.14in}{1.2in}{1.14in}{0pt}{0mm}{0pt}{0.4in}
\begin{document}

\begin{center}
\Large \bf
Bayesian Regression Analysis of Data with Random Effects Covariates from Nonlinear Longitudinal 
Measurements
\end{center}

\begin{center}
 \large Rolando De la Cruz$^{1,2,\star}$, Cristian Meza$^{3,\star}$, Ana Arribas--Gil$^{4,\star}$  and \\ Raymond J. Carroll$^{5,\star}$\\
 {\small $^{1}$Department of Public Health, School of Medicine,
 Pontificia Universidad Cat\'olica de Chile, \\ Marcoleta 434, Casilla 114D,
 Santiago, CHILE. \\
 $^{2}$Department of Statistics, Faculty of Mathematics, 
 Pontificia Universidad Cat\'olica de Chile, \\ Casilla 306, Correo 22,
 Santiago, CHILE. \\
 $^{3}$Centro de Investigaci\'on y Modelamiento de Fen\'omenos Aleatorios -- CIMFAV, \\
Faculty of Engineering,  Universidad de Valpara\'{\i}so, 
Av. Pedro Montt 2412, Valpara\'{\i}so, CHILE. \\
 $^{4}$ Departamento de Estad\'istica, Universidad Carlos III de Madrid, Getafe, Spain\\
$^{5}$ Department of Statistics, Texas A\&M University, College Station, Texas 77843 USA.\\
 $^{\star}$Email: \texttt{rolando@med.puc.cl} \,\,\, \texttt{cristian.meza@uv.cl}\,\,\, \texttt{ana.arribas@uc3m.es}\,\,\, \texttt{carroll@stat.tamu.edu}

}
\end{center}

\begin{center}
{\bf Abstract}
\end{center}
Joint models for a wide class of response variables and longitudinal measurements consist on a mixed--effects model to fit longitudinal trajectories whose random effects enter as covariates in a generalized linear model for the primary response. They provide a useful way to asses association between these two kinds of data, which in clinical studies are often collected jointly on a series of individuals and may help understanding, for instance, the mechanisms of recovery of a certain disease or the efficacy of a given therapy. The most common joint model in this framework is based on a linear mixed model for the longitudinal data. However, for complex datasets the linearity assumption may be too restrictive. Some works have considered generalizing this setting with the use of a nonlinear mixed--effects model for the longitudinal trajectories but the proposed estimation procedures based on likelihood approximations have been shown \citep{DelaCruz.etal.11} to exhibit some computational efficiency problems. In this article we propose an MCMC--based estimation procedure in the joint model with a nonlinear mixed--effects model for the longitudinal data and a generalized linear model for the primary response. Moreover, we consider that the errors in the longitudinal model may be correlated. We apply our method to the analysis of hormone levels measured at the early stages of pregnancy that can be used to predict \emph{normal} versus \emph{abnormal} pregnancy outcomes. We also conduct a simulation study to asses the importance of modelling correlated errors and quantify the consequences of model misspecification.




\section{Introduction}\label{sec:introd}
\begin{figure}
\centering\includegraphics[width=14cm]{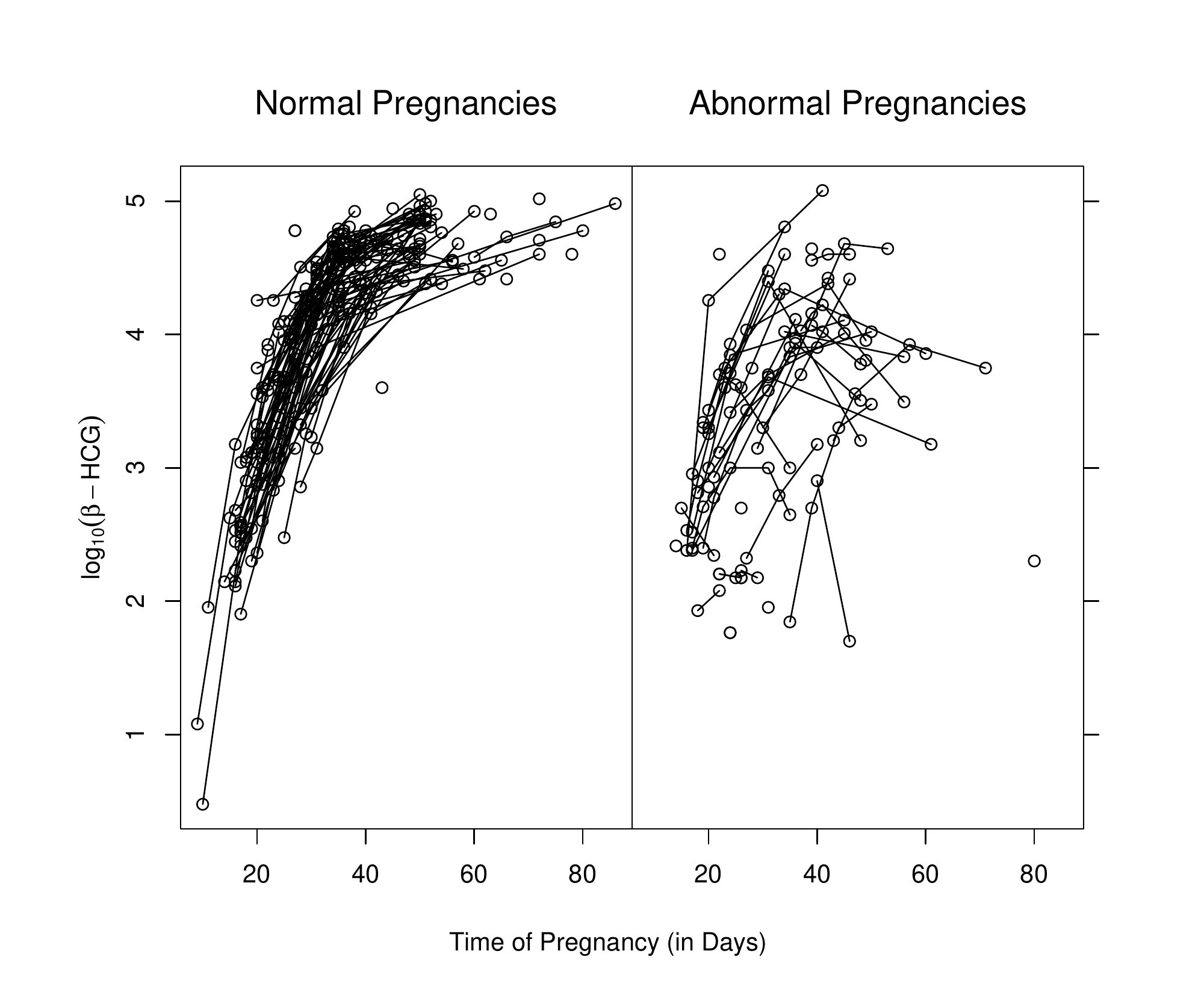}
\caption{\label{profiles} Observed $\beta$-HCG time profiles in the log scale for women with normal and abnormal pregnancy outcomes.}
\end{figure}

In many biomedical studies longitudinal biomarker profiles carry important information about the outcome of a therapy, a disease or a particular condition. In such cases, the association between the response or outcome and a series of longitudinal measurements is of primary interest. In Figure \ref{profiles} we illustrate one example that motivates the current paper. The longitudinal measurements of this dataset represent beta human chorionic gonadotropin ($\beta$-HCG) levels measured over time on 173 pregnant woman during the first 80 days of gestation. Here, the response of interest for each woman is given by her pregnancy outcome: \emph{normal}, if she had a normal delivery or \emph{abnormal} if she had any complication resulting in a nonterminal delivery and loss of the fetus. In such a framework a relevant question is how the variation of hormone concentration during the early stages of pregnancy may affect its outcome. In this case we are interested in a binary outcome but in a general setting we may be dealing with any kind of response.

If we observed longitudinal measurements without noise on a dense grid of time points this problem could be addressed from a functional perspective by using a logistic functional regression model with functional predictor and scalar response \citep{Ratcliffe02,Escabias04} or, more generally, a generalized functional linear model \citep{James02,Muller05}. However, this is an unrealistic setting in many biometrical applications in which the design for longitudinal data is irregular and sparse with very few observations available per individual and measurements are subject to experimental error. This is for instance the case in the $\beta$-HCG dataset in which the number of observations per women varies from 1 to 6, with a median of 2. 

Therefore, when dealing with noisy and highly sparse longitudinal trajectories a natural way of measuring their impact on the response of interest consists on extracting relevant latent information that could be used as covariates of a generalized linear model. Several authors have studied this problem focusing mainly on two types of response: binary outcomes and survival data.
\cite{Wang.etal.00} provided the first attempt in this direction with a joint model for longitudinal measurements and binary endpoints. They proposed to fit the longitudinal data with a linear mixed--effects model (LME) whose random effects were also covariates in a generalized linear model (GLM) for the binary endpoint.
The naive or two-step estimation method in such framework consists in fitting the LME and pluging-in the ordinary least squares estimates of the random effects in the GLM as if they were observed data. \cite{Wang.etal.00} showed that  this procedure introduces bias on the parameter estimates of the GLM and proposed several alternative approaches that reduced the bias. One of them is based on regression calibration, which in this context involves replacing the random effects by their estimated best linear unbiased predictors (BLUP) obtained by separately fitting the LME. Another strategy  relies on the use of pseudo-expected estimating equations (EEE).
For the same joint model \cite{Li.etal.04} relaxed the normality assumption of the random effects in the LME and provided estimators of the GLM parameters that yield consistency regardless of the true distribution. Furthermore, \cite{Li.etal.07} developed semiparametric likelihood-based inference for the GLM parameters and the random effects density. Recently, \cite{Horrocks09} used the \cite{Wang.etal.00} model to predict the achievement of successful pregnancy based on certain longitudinal measurements during a treatment for infertility. They estimated parameters using a Bayesian methodology similar to that proposed by \cite{GuoCarlin04} in the context of joint models for longitudinal and survival data. In that work, the focus was on predicting, from longitudinal measurements, the time to an event of interest instead of a binary outcome. The standard approach to tackle this question is again to fit a mixed--effects model to the longitudinal data whose random effects are covariates in a GLM for the time to event, see \cite{Neuhaus} for an overview.

For the pregnancy dataset that motivates this work it has been observed that log $\beta$-HCG levels and gestational age interact in a nonlinear way \citep{MarshallBaronSIM2000,DelaCruzQuintana07,DelaCruzQuintanaMuller07}, which suggests that a LME for longitudinal data may be inadequate in this case. Indeed, for the analysis of this dataset \cite{DelaCruz.etal.11} proposed a joint model in which the covariates for a primary logistic regression are the random effects of a nonlinear mixed--effects model (NLME) for hormone profiles. The authors compared several estimation methods including the naive two-step approach, BLUP and likelihood approximation methods based on several numerical integration techniques. They verified that as in the LME--GLM joint model, the first two procedures yield biased estimates. The third method seemed to work better for some particular approximation techniques, namely Laplacian and adaptive Gaussian approximations. However, these methods can be computationally inefficient in practice. \cite{WuHuWu08} also considered the problem of joint likelihood inference in the NLME-GLM model, although focusing on the case in which the primary outcome is the time to a given event, and encountered similar implementation problems. \cite{WuLiuHu10} proposed a fast and accurate joint estimation procedure for that model relying on the Laplace approximation. However, considering the findings of \cite{Joe08} about the asymptotic bias of estimators based on Laplace approximation for GLM with discrete response, these authors acknowledged that the performance of their method might be less satisfactory when dealing with binary outcomes instead of survival data.

To overcome these drawbacks, in this article we propose a Bayesian estimation approach for the NLME--GLM joint model. Although in its application to the pregnancy dataset we focus in the prediction binary outcomes, the general estimation framework that we describe is flexible enough to be used with any kind of response of interest.
Moreover, motivated by our real dataset, we assume that we may have autocorrelated error terms in the NLME. 

The rest of the paper is organised as follows. In Section \ref{sec:joint.model} we present the detailed specifications of the proposed joint model. In Section \ref{sec:estimation} we describe the MCMC algorithm for Bayesian estimation. A model comparison strategy is discussed in Section \ref{sec:CPO} and in Section \ref{sec:example} we apply our method to the $\beta$-HCG dataset. We compare the results to previous analyses on this dataset. In Section \ref{sim} we conduct a simulation study to asses the importance of model misspecification in the presence of autocorrelated errors. Finally, we offer a general discussion in Section \ref{disc}.

\section{Joint Model}\label{sec:joint.model}

The structure of interest here can be described by two components. The 
first component contains repeated observed measurements that are 
assumed to follow a nonlinear mixed--effects model over possibly 
unequally spaced times. The second component contains the primary 
outcome, which is assumed to follow a generalized linear model
where the random coefficients of the nonlinear mixed--effects models
are used as covariates.

Denote by $y_{ij}$, $i=1,\dots,m,\,\,j=1,\dots,n_i$, the observation of a 
continuous response for individual $i$ at time $t_{ij}$. Let $y_i=(y_{i1},y_{i2},\dots,y_{in_i})^{\prime}$ be the observed vector
of longitudinal measurement data at times $t_i=(t_{i1},t_{i2},\dots,t_{in_i})^{\prime}$. Assume that $y_i$ follows the nonlinear mixed--effects model 
\begin{equation}\label{eq:nlme}
 y_{i} = g(\alpha, X_i; t_{i}) + \epsilon_{i}, \qquad i=1,\dots,m,
\end{equation}
where $\alpha$ is a vector of $p$ unknown fixed effects parameters,
$X_i$ is a vector of $q$ unobservable random effects, $g$ is a real--valued nonlinear function of the 
fixed and random effects, and $\epsilon_{i}=(\epsilon_{i1},\dots,\epsilon_{in_i})^\prime$ is the within 
individual random error vector. We assume that the random effects $X_i$'s are independent 
and identically normally distributed with mean vector $\mu_{X}$ and covariance 
matrix $\Sigma_{X}$. Typically, the error terms $\epsilon_{i}$'s are assumed to be normal with zero mean vector and covariance matrix $\Sigma_{\epsilon_i}=\sigma^2_{\epsilon}I_{n_i}$, i.e. independent measurements errors, where $I_a$ denotes the identity matrix of dimension $a$.  However, in longitudinal data, measurements taken over time on individuals usually show a highly unbalanced structure (i.e. measurement times may be unequally spaced within an individual and may differ across individuals) and may be serially related. To take this into account we assume $\Sigma_{\epsilon_i}=\Sigma_{\epsilon_i}(\sigma^2_{\epsilon},\rho)$, with $\sigma^2_{\epsilon}$ being a scalar parameter and $\rho$ a vector of parameters describing the correlation structure. Depending on the context, various assumptions about the matrix $\Sigma_{\epsilon_i}(\sigma^2_{\epsilon},\rho)$  can be made 
\citep[see][Chap. 7]{VoneshChinchilli.BOOK}.
In the following we consider that $\Sigma_{\epsilon_i}(\sigma^2_{\epsilon},\rho)=\sigma^2_{\epsilon}\Sigma_i(\rho)$, where $\Sigma_i(\rho)$ is an $n_i \times n_i$ scaled matrix with $(k_1,k_2)$th element equal to $\rho^{|t_{ik_1}-t_{ik_2}|}$ though other choices are possible.  This matrix has a continuous time first-order autoregressive, CAR(1), structure \citep[see][]{delaCruz&Marshall:06}, which can cope with nonequally spaced measurements. We also assume that the $X_i$'s and $\epsilon_{i}$'s are mutually independent.

Now, assume that in addition to the $n_i$-dimensional vector 
of longitudinal measurements $y_i$, a primary response $D_i$,
and a $k$-vector of observed covariates, $W_i$, are observed on the $i$th individual.
We assume that the primary response and the random effects covariates are related via a GLM in canonical
form; i.e., the conditional distribution of $D_i$ given $X_i$ (and
$W_i$; conditioning on $W_i$ is dropped throughout) is
\begin{equation}\label{eq:glm}
 f(D_i | X_i; \theta) = \exp \left\{ 
 \frac{D_i(\beta_0^{\prime}W_i +\beta_1^{\prime}X_i)-b(\beta_0^{\prime}W_i +\beta_1^{\prime}X_i)}{a(\phi)} 
 + c(D_i,\phi) \right\},
\end{equation}
where $\theta=(\beta,\phi)^{\prime}$, with $\beta=(\beta_0^{\prime},\beta_1^{\prime})$,
are the parameters of primary interest; $\beta_0$ and $\beta_1$
are regression parameters, $\phi$ is a dispersion
parameter and $a(\cdot)$, $b(\cdot)$, $c(\cdot, \cdot)$ are known functions. In our context,
$\beta_1$ is of particular interest because it represents the relationship between
the primary response and features of longitudinal profiles.
As discussed in \cite{Wang.etal.00}, we can further assume that $y_i$ and $D_i$ are 
conditionally independent given $X_i$, in which case 
$$
f(y_i,D_i,X_i)=f(y_i,D_i|X_i)f(X_i)=f(y_i|X_i)f(D_i|X_i)f(X_i).
$$
The likelihood for the {\em joint} model $(y_i,D_i)$ is given by
\begin{equation}\label{eq:likelihood}
f(y,D) = \prod_{i=1}^m \int_{X} f(y_i|X_i)f(D_i|X_i)f(X_i)d\,X_i,
\end{equation}
where $y=(y_1,y_2,\dots,y_m)$ and $D=(D_1,D_2,\dots,D_m)$.
Note that the {\em joint} model $(y_i,D_i)$ is nonlinear in $X_i$, 
thus the integral in (\ref{eq:likelihood}) does not have a closed--form 
expression. However, approximation methods can be used to help the 
estimation. \cite{DelaCruz.etal.11} discuss methods based
on numerical integration techniques to obtain the MLE of the {\em joint model}
in the special case for which the primary response is binary.
In this paper we propose to estimate the model parameters using MCMC methods.


\section{Estimation via MCMC Methods}\label{sec:estimation}

Bayesian fitting of the {\em joint model} described in Section~\ref{sec:joint.model}
involves, as usual in the Bayesian framework, the updating from prior to posterior distributions for the parameters via appropriate likelihood functions. However, closed--form exact expressions for most of the relevant joint and marginal posterior distributions
are not available. Instead, we rely here on sampling-based approximations to the distributions of interest via Markov chain Monte Carlo (MCMC) methods: we use a Gibbs sampler or a Metropolis--within--Gibbs algorithm to explore the posterior.

We now consider the problem of choosing prior information for the parameters 
$\beta$, $\alpha$, $\mu_{X}$, $\Sigma_{X}$, $\sigma^2_{\epsilon}$, $\rho$, and 
$\phi$ of the {\em joint model}. We assume prior independence for parameters and
$$
\alpha \sim N_p(a_1,A), \quad \mu_X \sim N_q(c_1,C), \quad \Sigma_X \sim IW(v,vV), \quad \sigma^2_{\epsilon} \sim IG(v_1,v_2), 
$$

\begin{equation}\label{prior}
\rho \sim \pi(\rho), \quad \beta \sim N_r(s,S), \quad \textrm{and} \quad \phi \sim \pi(\phi).
\end{equation}

Here $IG(h, l)$ denotes the inverse gamma distribution, with shape parameter $h$ and scale parameter $l$, and
mean $(h-1)^{-1}l^{-1}$. By $V\sim IW(d,D)$, we mean that the random matrix $V$ follows an inverse Wishart distribution with scalar parameter $d$ and matrix parameter $D$ (by letting $V \sim IW(d,dD)$ we ensure that the mean of $V^{-1}$ equals $D^{-1}$). Also, $N_p(\mu,\Sigma)$ represents the $p$-variate normal distribution with vector mean $\mu$ and covariance matrix $\Sigma$, and $\pi(\cdot)$ stands for a general prior distribution to be specified in each case, as we discuss below.

In (\ref{prior}) the hyperparameters $(a_1,A,c_1,C,v,V,v_1,v_2,s,S)$, and those involved in the prior for $\rho$ and $\phi$, are all assumed to be known and chosen so that the priors are proper. In practice the specification of hyperparameters may be difficult, so
we can take the values of hyperparameters in such a way that we get non--informative priors
in the limiting case when no (or minimal) prior information is available. 

Note that in (\ref{eq:glm}), for binomial and Poisson primary responses, the dispersion parameter is $\phi=1$. In that case no prior specification is required for $\phi$ in (\ref{prior}). For normal primary response , $\phi$ is $\sigma^2$, and we can follow common practice in choosing an inverse gamma prior, $IG(r_1,r_2)$, for $\sigma^2$, i.e. $\pi(\sigma^2)=IG(r_1,r_2)$. In (\ref{prior}) we assume a uniform prior for $\rho$.

We now present the posterior density associated with the joint model. We will
note $f_N$, $f_{IG}$, $f_U$ and $f_{IW}$ the multivariate normal,
inverse gamma, uniform and inverse Wishart densities, respectively. Furthermore, $f_{GLM}$ denotes the primary response in the generalized linear model (\ref{eq:glm}). The joint posterior density of $X$, $\beta$, $\alpha$, $\mu_X$, $\Sigma_X$, $\sigma^2_{\epsilon}$, $\rho$, and 
$\phi$ given the observed data $d_m=\{ (y_i,D_i) \}_{i=1}^m$ is
\begin{equation}\label{posterior}
\pi(X,\beta,\alpha,\mu_X, \Sigma_X,\sigma^2_{\epsilon},\rho,\phi|d_m)=\frac{\pi^{*}(X,\beta,\alpha,\mu_X, \Sigma_X,\sigma^2_{\epsilon},\rho,\phi;d_m)}{m^{*}(d_m)}, 
\end{equation}
where the unnormalized posterior density is
\begin{align*}
\pi^{*}(X,\beta,\alpha,\mu_X, \Sigma_X,\sigma^2_{\epsilon},\rho,\phi;d_m) &= 
\left[\prod_{i=1}^m 
f_N(y_i;g(\alpha,X_i;t_i),\sigma^2_{\epsilon}\Sigma_i(\rho))f_{GLM}(D_i;X_i,\theta)f_N(X_i;\mu_X,\Sigma_X)\right] \\
& \qquad \times f_N(\alpha;a_1,A)f_{IG}(\sigma^2_{\epsilon};v_1,v_2)f_U(\rho)f_N(\mu_X;c_1,C)f_{IW}(\Sigma_X;v,vV)\\
& \quad \times f_N(\beta,s;S) \pi(\phi)
\end{align*}
and the normalizing constant (which is also the marginal density of the data) is
$$
m^{*}(d_m)= \int \pi^{*}(X,\beta,\alpha,\mu_X, \Sigma_X,\sigma^2_{\epsilon},\rho,\phi;d_m) 
dX\,d\beta\,d\alpha\,d\mu_X\,d\Sigma_X\,d\sigma^2_{\epsilon}\,d\rho\,d\phi.
$$
The full conditionals to implement the MCMC procedure can be easily derived from (\ref{posterior}). Indeed, we have 
\begin{align}
\label{Xpost}
\pi(X|\textrm{rest},d_m)& =\prod_{i=1}^m \pi(X_i|\textrm{rest},d_m), \\
\label{alpost}
\pi(\alpha|\textrm{rest},d_m) & \propto \pi(\alpha) \prod_{i=1}^m f(y_i|X_i)  \nonumber \\
& \propto\exp\left\{-\frac{1}{2} \textrm{tr}\left( \frac{1}{\sigma^2_{\epsilon}} \Sigma^{-1}(\rho)(y_i-g(\alpha,X_i;t_i))^{\prime}(y_i-g(\alpha,X_i;t_i))+ A^{-1}(\alpha-a)^\prime (\alpha-a) \right) \right\}, \\
\label{bepost}
\pi(\beta|\textrm{rest},d_m) & 
\propto\exp\left\{-\frac{1}{2} \textrm{tr}( S^{-1}(\beta-s)^\prime (\beta-s) + \sum_{i=1}^{n} \frac{D_i\theta_i-b(\theta_i)}{a(\phi)}  \right\}, \\
\label{mupost}
\pi(\mu_{X}|\textrm{rest},d_m) & \propto  \pi(\mu_{X}) \prod_{i=1}^m f(X_i),  \\
\label{sigpost}
\pi(\Sigma_{X}|\textrm{rest},d_m) & \propto  \pi(\Sigma_{X}) \prod_{i=1}^m f(X_i), \\
\label{sig2post}
\pi(\sigma^2_{\epsilon}|\textrm{rest},d_m) & \propto \pi(\sigma^2_{\epsilon})
\prod_{i=1}^m f(y_i|X_i), \\
\label{ropost}
\pi(\rho|\textrm{rest},d_m) & \propto \pi(\rho)\prod_{i=1}^m f(y_i|X_i) \nonumber \\
& \propto \exp\left\{-\frac{1}{2} \textrm{tr}\left( \frac{1}{\sigma^2_{\epsilon}} 
\Sigma^{-1}(\rho)(y_i-g(\alpha,X_i;t_i))^{\prime}(y_i-g(\alpha,X_i;t_i))\right) \right\},\\
\label{phipost}
\pi(\phi|\textrm{rest},d_m) & \propto \pi(\phi)\prod_{i=1}^m f(D_i|X_i),
\end{align}
where $\theta_i = \beta_0^{\prime}W_i+\beta_1^{\prime}X_i$ and $rest$ denotes the remaining components of the model to which we are conditioning in each case. Some of these densities have a closed-form expression. Indeed, from (\ref{mupost}), (\ref{sigpost})  and  (\ref{sig2post}) it is easy to check that $\mu_{X}|\textrm{rest},d_m$ is multivariate normal with mean
$$
(m\Sigma_X^{-1}+C^{-1})^{-1}(\Sigma_X^{-1}\sum_{i=1}^m X_i+c_1C^{-1})
$$
and covariance matrix $(m\Sigma_X^{-1}+C^{-1})^{-1}$. Also, $\Sigma_{X}|\textrm{rest},d_m$ follows an inverse Wishart distribution with scale parameter $v+\sum_{i=1}^m n_i$ and matrix parameter
$$
vV+\sum_{i=1}^m (y_i-g(\alpha,X_i;t_i))^{\prime}(y_i-g(\alpha,X_i;t_i)).
$$
Finally, $\sigma^2_{\epsilon}|\textrm{rest},d_m$ follows an inverse gamma distribution with shape parameter $N/2+v_1$ and scale parameter
$$
\left(\frac{1}{v_2} + \frac{\sum_{i=1}^m RSS_{y_i}}{2} \right)^{-1},
$$
where $RSS_{y_i}=(y_i-g(\alpha,X_i;t_i))^{\prime}\Sigma^{-1}_i(\rho)(y_i-g(\alpha,X_i;t_i))$.
Due to the fact that $g(\cdot)$ is a nonlinear function of $X_i$, the full conditional density in (\ref{Xpost}), $\pi(X_i|\textrm{rest},d_m)$, cannot be written explicitly.
However, the full conditional density of $X_i$ can be written, up to a constant of
proportionality, as
$$
\exp \left \{-\frac{1}{2} \textrm{tr} \left( \frac{1}{\sigma^2_{\epsilon}} 
\Sigma^{-1}(\rho)(y_i-g(\alpha,X_i;t_i))^{\prime}(y_i-g(\alpha,X_i;t_i)) 
+ \Sigma_X^{-1}(X_i-\mu_X)^{\prime}(X_i-\mu_X) \right) \right) 
$$
\begin{equation}\label{eq:Xi.prop}
\left. + \frac{D_i(\beta_0^{\prime}W_i+\beta_1^{\prime}X_i)-b(\beta_0^{\prime}W_i+\beta_1^{\prime}X_i)}{a(\phi)}\right\}. \\
\end{equation}
In this case, to simulate from this full conditional we use a Metropolis--Hastings algorithm
within each Gibbs step. Because (\ref{eq:Xi.prop}) is known up to
a normalization constant, we can compute its mode $X_i^{\star}$ and
Hessian $V_i^{\star}$ using numerical optimization techniques. This
yields a natural choice of the proposal distribution, 
a multivariate normal distribution with mean vector
$X_i^{\star}$ and variance--covariance matrix $V_i^{\star -1}$,
denoted by $f_N(X_i;X_i^{\star},V_i^{\star -1})$. Then
we can implement the Metropolis--Hastings algorithm
as follows. Denote $X_i^{(r)}$ the current value
of $X_i$ at the $r$th iteration. A new candidate value
$X_i^{c}$ is drawn from the proposal distribution $f_N(X_i;X_i^{\star},V_i^{\star -1})$.
The acceptance probability is computed as:
$$
\min\left\{1,\frac{f_N(X_i^{c};X_i^{\star},V_i^{\star -1})}{f_N(X_i^{(r)};X_i^{\star},V_i^{\star -1})}
\frac{f_N(y_i;g(\alpha,X_i^{c};t_i),\sigma_{\varepsilon}^{2}\Sigma_i(\rho))f_{GLM}(D_i;X_i^{c},\theta)
f_N(X_i^{c};\mu_X,\Sigma_X)}{f_N(y_i;g(\alpha,X_i^{(r)};t_i),\sigma_{\varepsilon}^{2}\Sigma_i(\rho))f_{GLM}(D_i;X_i^{(r)},\theta)
f_N(X_i^{(r)};\mu_X,\Sigma_X)}  \right\}.
$$
Note that there is no need to compute the normalization constant because it cancels
out in the acceptance probability. For the remaining full conditionals, no such closed--form expression exists either and the same Metropolis--Hastings within Gibbs algorithm is used to obtain draws from them. Note that the full conditional of the dispersion parameter $\phi$ of the GLM is only required depending on the kind of the primary response. For instance, for the binomial and Poisson model we have $\phi=1$.

The Markov chain associated with the MCMC algorithm is denoted by 
$\Phi \{( X^{(n)},\alpha^{(n)},\beta^{(n)},$ $\mu_X^{(n)},\Sigma_X^{(n)},\sigma^{2(n)}_{\epsilon},\rho^{(n)},\phi^{(n)})\}_{n=0}^{\infty}$ and has the posterior density (\ref{posterior}) as its stationary density. To run the algorithm, given the current state,
$( X^{(n)},\alpha^{(n)},\beta^{(n)},\mu_X^{(n)},\Sigma_X^{(n)},\sigma^{2(n)}_{\epsilon},\rho^{(n)},\phi^{(n)})$, we draw each of the $X_i^{(n+1)}$'s independently
and form $X^{(n+1)}$. Then, the following series of steps is conducted: given $X^{(n+1)}$, $\sigma^{2(n)}_{\epsilon}$ and $\rho^{(n)}$,
we draw $\alpha^{(n+1)}$; given $X^{(n+1)}$ and $\phi^{(n)}$ we draw $\beta^{(n+1)}$;
given $X^{(n+1)}$ and $\Sigma_X^{(n)}$ we draw $\mu_X^{(n+1)}$; given $X^{(n+1)}$, $\alpha^{(n+1)}$ and $\mu_X^{(n+1)}$ we draw $\Sigma_X^{(n+1)}$;
given $X^{(n+1)}$, $\alpha^{(n+1)}$ and $\rho^{(n)}$ we draw $\sigma_{\epsilon}^{2(n+1)}$; given $X^{(n+1)}$ and $\alpha^{(n+1)}$ we draw $\rho^{(n+1)}$; and finally, given
$X^{(n+1)}$ and $\beta^{(n+1)}$ we draw $\phi^{(n+1)}$.


\section{Model Comparison}\label{sec:CPO}
The conditional predictive ordinate (CPO) statistics introduced by \cite{Gelfand:92} is a popular and useful model assessment tool based on the marginal posterior predictive density of the response for individual $i$ given the observed data from the rest of the individuals. Let $\theta=(\beta,\alpha,\mu_X,\Sigma_X,\sigma^2_{\epsilon},\rho,\phi)$ be the parameters of the {\em joint model}, let $d_m$ be the observed data for all individuals, and let $d_{-(i)}$ and $X_{-(i)}$ denote the observed data and random--effects vector, respectively, of the whole sample excluding individual $i$. Further, let us note $d_i=(y_i,D_i)$ where, for individual $i$, $y_i$ is the observed vector of longitudinal measurements and $D_i$ is the primary response of the GLM. Then, the $CPO$ statistic for individual $i$ for our {\em joint model} is defined as
\begin{align*}
CPO_i & = f(d_{-(i)}|d_m) \\
& = \left [ \frac{f(d_{-(i)})}{f(d_m)} \right]^{-1} \\
& = \left [ E_{\theta,X|d_m} \left ( \frac{1}{f(y_i|X_i,\theta)f(D_i|X_i,\theta)f(X_i|\theta)} \right) \right]^{-1}.
\end{align*}
A Monte Carlo estimate of $CPO_i$ can be obtained by using a single MCMC sample from the posterior distribution
$\pi(X,\theta|d_m)$. Let $(\theta^{(1)},X^{(1)}_i),\dots,(\theta^{(R)},X^{(R)}_i)$ be a sample of size $R$, for corresponding parameters
and individual--specific random effect, drawn from $\pi(\theta,X|d_m)$ after the burn-in phase. A natural Monte Carlo approximation of $CPO_i$ is given by
$$
\widehat{CPO}_i \approx \left [ \frac{1}{R} \sum_{r=1}^R\frac{1}{f(y_i|X_i^{(r)},\theta)f(D_i|X_i^{(r)},\theta^{(r)})f(X_i^{(r)}|\theta^{(r)})} \right]^{-1} .
$$
For each individual, larger values of CPO imply a better fit of the model. As a summary statistic of CPO over all individuals, we use the logarithm of the pseudomarginal likelihood (LPML; \citealp{Ibrahim:01}), which is defined by
\begin{equation}\label{LPML}
LPML = \frac{1}{m} \sum_{i=1}^m \log \left( \widehat{CPO}_i \right).
\end{equation}


\section{Analysis of Pregnant Women Data}\label{sec:example}

The main objective of the analysis of the pregnant women dataset presented
in Section~\ref{sec:introd} is to investigate the effects of the 
$\beta$--HCG longitudinal process on pregnancy outcomes,  and in
particular the association between normal pregnancy and features of 
longitudinal $\beta$--HCG profiles. The data were collected 
from a total of $173$ young pregnant women over a period of $2$ years in a private fertilization obstetrics clinic 
in Santiago, Chile. The resulting dataset consists of $124$ patients 
whose pregnancies developed without any complications and $49$ patients with abnormal 
pregnancies. Let $D_i = 1$ and $0$ denote normal and abnormal 
pregnancy outcomes, respectively, for woman $i$, $i=1,\dots,m$, ($m=173$). For the longitudinal 
$\beta$--HCG concentrations, the $173$ women altogether contribute a 
total of $375$  observations, where the number of observations $n_i$ 
per woman ranges from 1 to 6 (median 2). Approximately $30\%$ of the 
$173$ women have only one $\beta$--HCG measurement, $31\%$ have two, $33\%$ 
have three, and only $6\%$ have four or more measurements. 

As discussed in previous work \citep{MarshallBaronSIM2000,DelaCruzQuintana07,DelaCruzQuintanaMuller07}, 
a reasonable representation of the log $\beta$--HCG profile ($y_i$) for the $i$th woman is 
\begin{equation}\label{eq:tplm}
y_{i} = \frac{X_{i}}{1+\exp\{-(t_{i}-\alpha_1)/\alpha_2 \}} + \epsilon_{i}
\end{equation}
where time is measured in days and the measurement errors $\epsilon_{i}$ are Gaussian. For this dataset, it seems reasonable to consider the error distribution $\epsilon_{i}\sim N_{n_i}(0,\Sigma_{\epsilon_i}(\sigma^2_{\epsilon},\rho) )$ where $\Sigma_{\epsilon} (\sigma^2_{\epsilon},\rho)$ 
is a correlation structure with unknown $\sigma^2_{\epsilon}$ and $\rho$ parameters. In particular, we consider the CAR(1) correlation structure described in Section \ref{sec:joint.model}. The woman--specific random effect $X_{i}$ is assumed to satisfy $X_{i} \sim N(\mu_{X},\sigma^2_X)$ and it represents the asymptotic behaviour of the log $\beta$--HCG profile. To describe 
the relation between the pregnancy outcome and $X_{i}$, we consider the 
primary logistic regression model 
\begin{equation}\label{eq:lm}
\Pr(D_i = 1|X_i) = [1+\exp\{ -(\beta_1 + \beta_2 X_{i}) \}]^{-1}.
\end{equation}
We used the Bayesian approach described in Section \ref{sec:estimation} to estimate the parameters of this joint model.
To illustrate the gain obtained by considering correlated errors, we also fitted the same joint model with independent errors in (\ref{eq:tplm}). We also considered separate fitting, i.e. we estimated independently the NLME (\ref{eq:tplm}) and the GLM (\ref{eq:lm}), assuming both independent and correlated errors.

Implementing Gibbs sampling requires adopting specific values for the hyperparameters ($a_1$, $A$, $c_1$, $C$, $v$, $V$, $v_1$, $v_2$, $s$, $S$). We considered weakly informative prior distributions for the parameters in all the models. The values for the hyperparameters were taken as follows: $a_1=s=(0,0)$, $A=S=1\,000I_2$, $c_1=0$, $C=1\,000$, $v=6$, $V=0.00083$, $v_1=3$ and $v_2=0.01$. We also performed the analysis with different hyperparameter values, obtaining very similar results. This suggests robustness to the hyperparameter choices. Always, the choice of the hyperparameters values was made to use diffuse proper priors. We performed $2\,000\,000$ iterations of the MCMC procedure. After the first $10\,000$ iterations, samples were collected, at a spacing of 50 iterations, to obtain approximately independent samples. We ended up with $R=39\,800$ samples to calculate posterior quantities of interest. The program used to fit the model was written in Fortran, but let us point out that the model for the i.i.d. case can be fitted in OpenBUGS. 
To diagnose convergence, we suggest any of the convergence criteria discussed in the literature, for example, those included in the BOA package \citep{Smith.04}. We prefer to use diagnostics which do not require multiple parallel chains, as proposed by \cite{Geweke.92}. In this analysis, applying Geweke's convergence criterion separately to each model parameter, where the absolute value of the z statistics was less than 1.6 in all cases, showed that convergence had been achieved.

Table~\ref{estimates} presents the results obtained by fitting the joint model 
(\ref{eq:tplm})-(\ref{eq:lm}) by the procedure described in this article and also the estimates provided by MCMC methods for the separate fitting. For both strategies, we considered independent and correlated errors for the NLME model. For each parameter and each model, the posterior mean, the standard error and the posterior median together with a 95\% credibility interval are given.\\

\begin{table}
\caption{\label{estimates} Parameter estimates for joint and separate modeling.}
\begin{center}
\begin{tabular}{crrrrrcrrrrr}\hline
           & \multicolumn{5}{c}{\it{Joint Model}} & &  \multicolumn{5}{c}{\it{Separate Model}} \\ \cline{2-6} \cline{8-12}       
           & 
\multicolumn{1}{c}{Mean} &   
\multicolumn{1}{c}{SD} & 
\multicolumn{1}{c}{2.5\%} &
\multicolumn{1}{c}{Median} & 
\multicolumn{1}{c}{97.5\%} &  &
\multicolumn{1}{c}{Mean} &
\multicolumn{1}{c}{SD} &
\multicolumn{1}{c}{2.5\%} &
\multicolumn{1}{c}{Median} &
\multicolumn{1}{c}{97.5\%} \\ \hline
\multicolumn{12}{l}{{\bf Independent Errors}} \\ \hline
\multicolumn{1}{c}{} &   
\multicolumn{11}{c}{Longitudinal submodel} \\
$\mu_{X}$ & 4.495 & 0.063 & 4.375 & 4.494 & 4.620 &  & 4.513 & 0.065 & 4.388 & 4.512 & 4.643  \\ 
$\alpha_1$ &14.850 & 0.400 &14.040 &14.870 &15.590 &  &15.000 & 0.392 &14.190 &15.020 &15.740  \\   
$\alpha_2$ & 7.467 & 0.520 & 6.510 & 7.446 & 8.551 &  & 7.482 & 0.527 & 6.515 & 7.461 & 8.581  \\
$\sigma^2_{\epsilon}$ & 0.132 & 0.014 & 0.108 & 0.131 & 0.161 & & 0.131 & 0.014 & 0.107 & 0.130 & 0.161 \\
$\sigma^2_{X}$ & 0.290 & 0.045 & 0.211 & 0.287 & 0.388 & & 0.294 & 0.047 & 0.212 & 0.291 & 0.395 \\ 
\multicolumn{1}{c}{} &   
\multicolumn{11}{c}{Logistic submodel} \\
$\beta_1$  & -15.280 & 3.957 & -24.340 & -14.850 & -8.788 &  & -14.460 & 2.868 & -20.450 & -14.320 & -9.224 \\
$\beta_2$  & 3.682 & 0.902 & 2.204 & 3.576 & 5.737 &  & 3.443 & 0.638 & 2.279 & 3.413 & 4.777 \\
\hline
\multicolumn{12}{l}{{\bf Correlated Errors}} \\ \hline
\multicolumn{1}{c}{} &   
\multicolumn{11}{c}{Longitudinal submodel} \\
$\mu_{X}$ & 4.495 & 0.063 & 4.373 & 4.494 & 4.621 &  & 4.521 & 0.064 & 4.399 & 4.519 & 4.649  \\ 
$\alpha_1$ &15.180 & 0.409 &14.340 &15.190 &15.940 &  &15.330 & 0.433 &14.460 &15.340 &16.160  \\   
$\alpha_2$ & 7.211 & 0.487 & 6.311 & 7.193 & 8.228 &  & 7.278 & 0.504 & 6.361 & 7.256 & 8.331  \\
$\sigma^2_{\epsilon}$ & 0.187 & 0.025 & 0.143 & 0.185 & 0.240 & & 0.250 & 0.053 & 0.162 & 0.245 & 0.359 \\
$\sigma^2_{X}$ & 0.223 & 0.046 & 0.141 & 0.220 & 0.322 & & 0.127 & 0.075 & 0.003 & 0.128 & 0.275 \\ 
$\rho$ & 0.924 & 0.017 & 0.884 & 0.927 & 0.951 && 0.944 & 0.017 & 0.903 & 0.947 & 0.968 \\
\multicolumn{1}{c}{} &   
\multicolumn{11}{c}{Logistic submodel} \\
$\beta_1$  & -22.860 & 5.474 & -34.790 & -22.400 & -13.400 &  & -39.040 & 6.965 & -53.530 & -38.730 & -26.450 \\
$\beta_2$  & 5.431 & 1.259& 3.261 & 5.325 & 8.174 &  & 8.885 & 1.546 & 6.088 &  8.815 & 12.110 \\
\hline
\end{tabular}
\end{center}
\end{table}

From Table~\ref{estimates}, we can see that there are no important differences between the parameter estimates obtained from joint and separate fitting under the assumption of independent errors. However, if we assume correlation in the error term, we obtain, as expected, a significant difference in the GLM parameter estimates $\beta_1$ and $\beta_2$ obtained from joint and separate fitting.

Now, from the estimated parameter values we get estimates of $P(D_i|X_i)$, which allows us to consider the underlying classification problem and compare the four models performances. To do so, we calculated the confusion matrix of classification which contains information about correspondence between actual and predicted classes. A probability cut-off value of 0.5 was considered as classification rule. The results are presented in Table \ref{tab:results.clas.1}. 


\begin{table}
\caption{Confusion matrix of classification 
for the joint and separate fitting with independent and correlated errors.}
\label{tab:results.clas.1}
\begin{center}
\begin{tabular}{ccccccccc}

&&&&&&&&\\\hline
&&&&&&&&\\
&& \multicolumn{2}{c}{\textit{Joint Model}} & &\multicolumn{2}{c}{\textit{Separate Model}} && \\
&&&&&&&&\\\cline{3-4}\cline{6-7}
&&&&&&&&\\
\textit{Group}&& \textit{Normal} & \textit{Abnormal} & & \textit{Normal} & \textit{Abnormal} && \textit{Total}\\
&&&&&&&&\\\hline
&&&&&&&&\\
&&\multicolumn{5}{c}{Independent Errors} &&\\
&&&&&&&&\\\cline{3-8}
&&&&&&&&\\
Normal && 122   & \,\,\,2     &   &    120   & \,\,\,4  & &  124 \\
Abnormal && \,\,\,21 &     28      &   & \,\,\,26 &     23   && \,\,\,49  \\
&&&&&&&&\\\hline
&&&&&&&&\\
\textit{Total}&& 143 & 30  && 146 &  27 && 173  \\
&&&&&&&&\\\hline
&&&&&&&&\\
&&\multicolumn{5}{c}{Correlated Errors} &&\\
&&&&&&&&\\\cline{3-8}
&&&&&&&&\\
Normal && 124 & 0  && 119 & \,\,\,5  & &  124\\
Abnormal && 13 & 36  &&  \,\,\,21 &     28   && \,\,\,49   \\
&&&&&&&&\\\hline
&&&&&&&&\\
\textit{Total}&& 137  & 36  && 140 & 33  & & 173 \\
&&&&&&&&\\\hline
\end{tabular}
\end{center}
\end{table}

\begin{table}
\caption{Error-rate, sensitivity, specificity and area under curve (AUC) for joint and separate models. In parenthesis, the standard deviation of AUC. }
\label{tab:results.summary}
\begin{center}
\begin{tabular}{cccccc}
&& Error-rate & Sensitivity & Specificity & AUC (s.d.) \\\hline
Joint Model:  &Errors&&&\\ \cline{2-2}
&Independent & 13.3\% & 98.4\% & 71.8\% & 0.908 (0.032)\\
&Correlated & 7.5\%  & 100\% & 73.5\% & 0.988 (0.007)  \\\hline
Separate Model: &Errors&&&\\ \cline{2-2}
& Independent & 17.3\% & 96.8\% & 46.9\%  & 0.792 (0.046)\\
&Correlated & 15.0\% & 96.0\% & 57.1\% & 0.815 (0.044) \\\hline
\end{tabular}
\end{center}
\end{table}

Table \ref{tab:results.summary} shows the error rate, the sensitivity, and the specificity of the classification rule with a probability cut-off value of 0.5 for the four models. It also presents the area under the Receiver Operating Characteristic (ROC) curve (AUC) and its standard deviation. The ROC curve represents the sensitivity versus 1 minus the specificity for any cut-off value from $0$ to $1$. Then, a larger value of AUC means a better classifying performance. In the case of independent errors, we found an 
error rate estimation of approximately $13.3\%$ and $17.3\%$ for the joint and 
separate models respectively. As discussed before by \cite{DelaCruz.etal.11}, the joint model seems to improve classification. Now, considering a CAR correlation structure in the errors, we obtained an 
error rate estimation of approximately $7.5\%$ and $15.0\%$ for the joint and 
separate models, respectively. Therefore, it is clear that the inclusion of correlation structure allows to significantly improve the classification results in this dataset. We observe the same kind of improvement for the sensitivity, the specificity and the AUC for the joint correlated model versus the other three models. It then appears evident that the joint strategy with correlation structure in the error term globally improves the sensitivity and the specificity for predicting a normal pregnancy outcome for this population of women.\\

To further compare the two joint models, this time in terms of fitting accuracy, we calculated for each one the $LPML$ (\ref{LPML}), as defined in Section \ref{sec:CPO}. Models with greater $LPML$ values will indicate a better fit. We found $LPML= -321.03$ for the joint model with correlated errors and $LPML= -350.26$ for the joint model assuming independent errors. This suggests that the joint model with a correlation structure in the errors provides a marginally better fit to this specific dataset.

We compare our results with those found using the Bayesian longitudinal discriminant analysis (BLDA) approach (see De la Cruz--Mes\'{\i}a and Quintana, 2007) in which case the reported error rate was approximately 16\% which is greater than under the joint model with correlated errors, 7.5\%. The same happens with the sensitivity and the specificity: with the BLDA approach the sensitivity was found to be 95\% and the specificity 57\%.


\section{Simulation Study}\label{sim}
To assess the importance of considering correlation in the error term of the NLME on synthetic data, we conducted the simulation study described below. The objective is to show the effect of misspecification regarding the error dependence structure.

We used the joint model (\ref{eq:tplm})-(\ref{eq:lm}) to simulate observations that replicate the sparse structure of the real dataset used in Section \ref{sec:example}. Indeed, we kept the same number of individuals in each group and for each individual, the same number of observations as well as the same observation time points. We simulated 500 datasets using the following parameter values:
\begin{eqnarray*}
\mu_X &=& 4\\
\alpha&=&(\alpha_l,l=1,2)=(15,7)\\
\beta&=&(\beta_h,h=1,2)=(-22,5)\\
\sigma^2_X&=&0.2\\
\sigma^2_{\epsilon}&=&0.2\\
\rho&=&0.9
\end{eqnarray*}
The generated datasets were analysed using the estimation procedure presented in Section \ref{sec:estimation} but considering that the error terms $\epsilon_i$ are independent, i.e.  $\Sigma_{\epsilon_i} (\sigma^2_{\epsilon},\rho)= \sigma^2_{\epsilon} I_{n_i}$. This strategy allows us to analyse the bias introduced by this misspecified model which does not consider the correlation structure of the data. We also compared the results obtained with those of a joint model with correlated errors.

\begin{figure}[!htp]
\begin{center}
  \includegraphics[width=15cm]{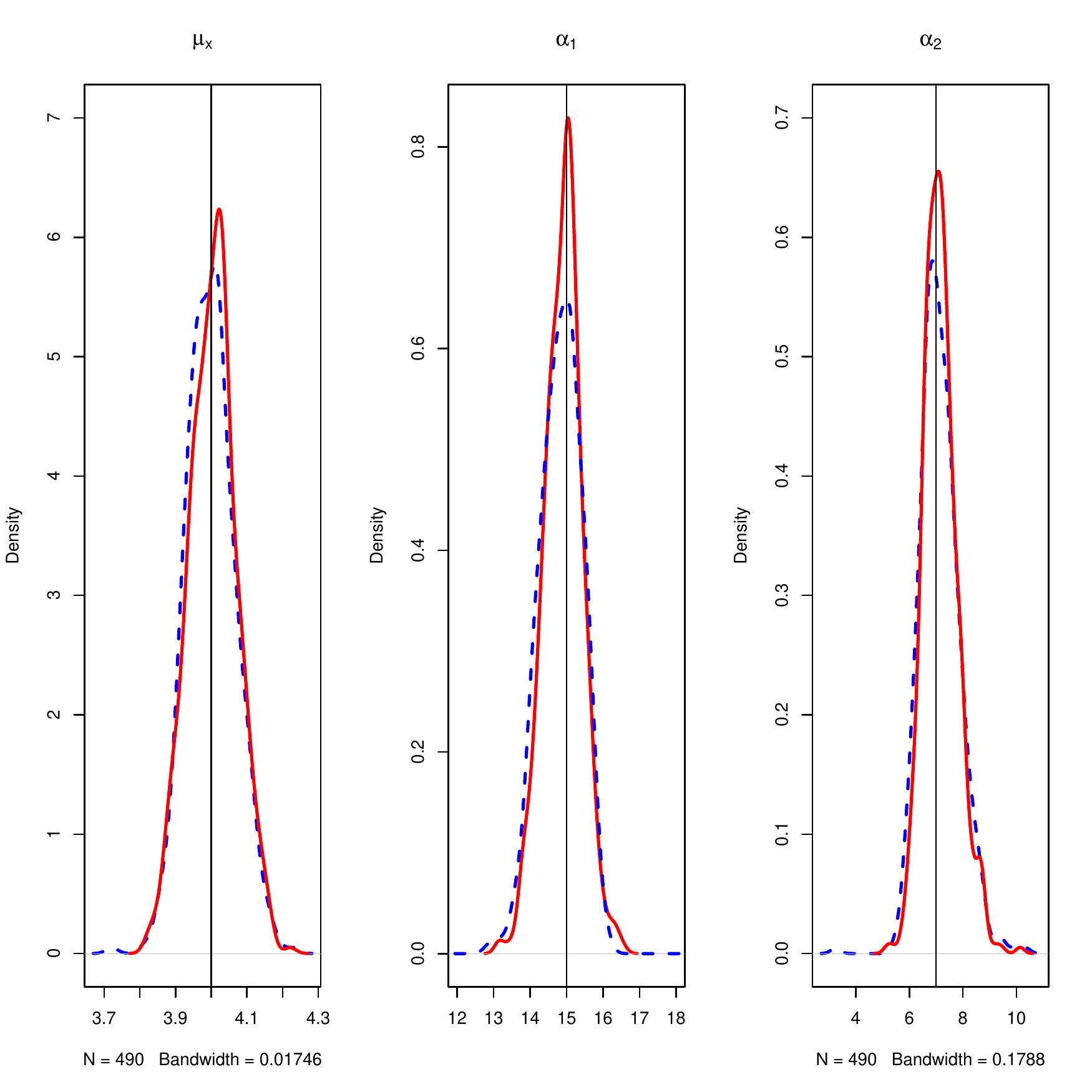}
  \caption{Longitudinal submodel: Distribution of fixed effects parameter estimates over 500 simulated datasets using a joint model with independent (dashed line) and correlated (solid line) errors.  Vertical lines represent true values.}\label{graf:nlme}
\end{center}
\end{figure}

\begin{figure}[!htp]
\begin{center}
  \includegraphics[width=15cm]{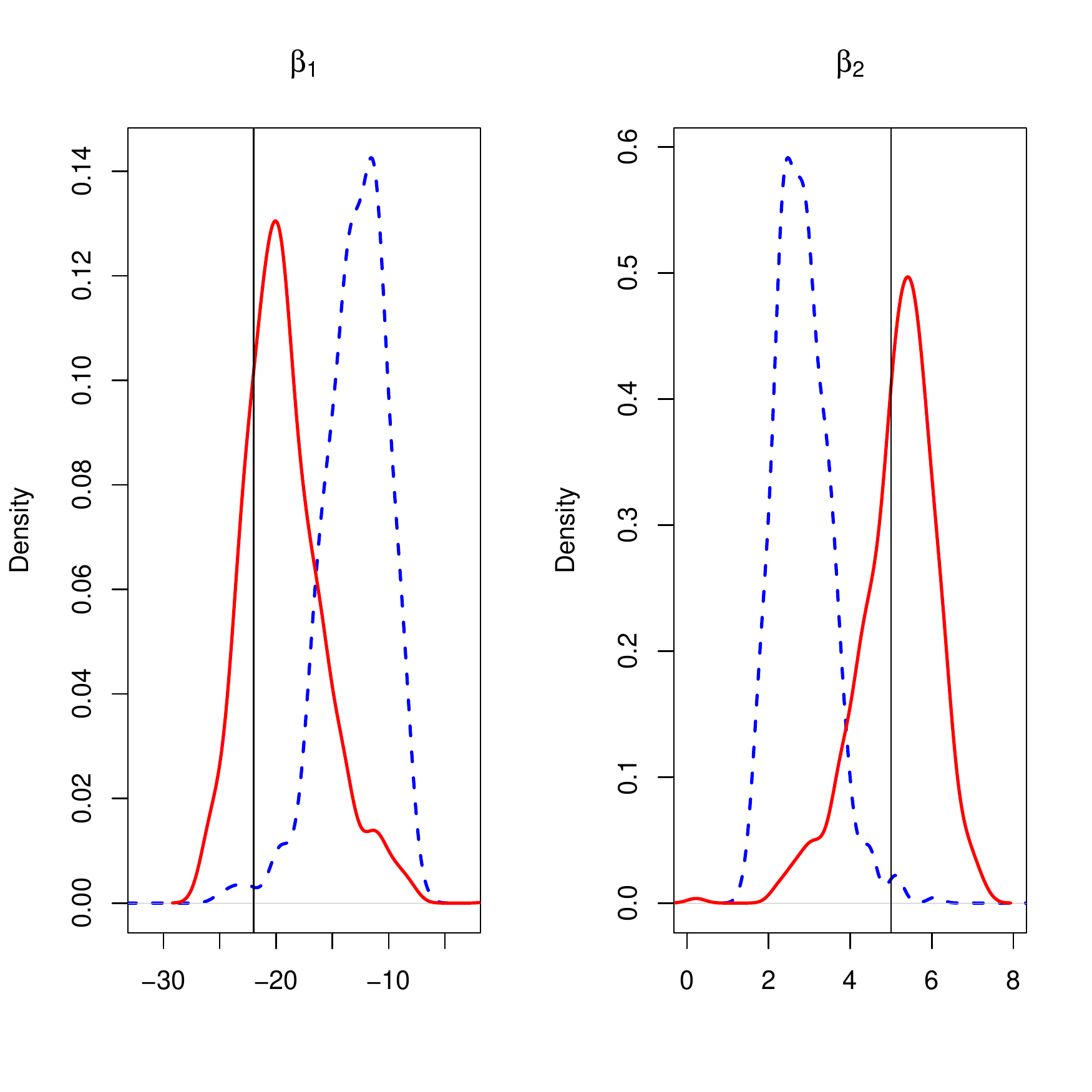}
  \caption{Logistic submodel: Distribution of fixed effects parameter estimates over 500 simulated datasets using a joint model with independent (dashed line) and correlated (solid line) errors. Vertical lines represent true values.}\label{graf:glm}
\end{center}
\end{figure}

Summary statistics for the Bayesian estimates obtained for these 500 simulated datasets are given in Table \ref{Table500}. The true values of the parameters used in the simulation, the means and the medians with their respective standard errors, and individual coverage probability are provided. It can be seen that the mean and median values for the logistic submodel parameters present important biases. Specifically,  when we use the misspecified model, we observe an important overestimation for $\beta_1$ and an underestimation for $\beta_2$. Instead, as expected, we get much better results when we consider correlated errors. For the real dataset, in Table \ref{estimates}, we observed a similar behaviour since for the joint model with correlated errors the estimate of $\beta_1$ decreased in almost 50\% in comparison with the estimate obtained under the independent error assumption whereas for $\beta_2$ we observed an increase of almost 50\%. On the contrary, the nonlinear model parameters estimates are very close to the simulated values for both models. We can observe the same behaviour in terms of coverage probabilities. Figures \ref{graf:nlme} and \ref{graf:glm} provide a graphical representation of these results displaying the distribution of estimates of the longitudinal and logistic submodel parameters. This simulation study shows that not taking into account correlation among errors in the longitudinal measurements of the joint model may introduce large bias in GLM parameter estimates.

\begin{table}[!htp]
\caption{Results obtained on 500 simulated datasets for a joint model with independent and correlated errors.}\label{Table500}
\begin{tabular}{ccccccc}\hline
& True Value & Mean & $SD_{\bar{X}}$ & Median & $SD_{Median}$ & Coverage Prob.\\\hline
\multicolumn{7}{l}{{\bf Independent Errors}} \\ \hline
{\it Longitudinal submodel} &&&&&&\\
$\mu_X$ & 4.00 & 3.998 & 0.067 & 3.998 & 0.065 & 0.95\\
$\alpha_1$ & 15 & 14.86 & 1.239 & 14.88 & 0.546 & 0.91 \\
$\alpha_2$ & 7 & 7.137 & 0.738 & 7.095 & 0.698 & 0.91 \\
$\sigma^2_{\epsilon}$ & 0.2 & 0.147 & 0.022 & 0.144 & 0.015 & 0.12\\
$\sigma^2_X$ & 0.2 & 0.285 & 0.044 & 0.282 & 0.044 & 0.44\\
$\rho$ & 0.9 & - & - & - & - & -\\
{\it Logistic submodel} &&&&&&\\
$\beta_1$ & -22 & -13.06 & 3.330 & -12.823 & 3.067 & 0.31\\
$\beta_2$ & 5 & 2.861 & 0.797 & 2.805 & 0.725 & 0.30\\\hline
\multicolumn{7}{l}{{\bf Correlated Errors}} \\ \hline
{\it Longitudinal submodel} &&&&&&\\
$\mu_X$ & 4.00 & 4.004 & 0.067 & 4.003 & 0.066 & 0.938\\
$\alpha_1$ & 15 & 14.90 & 0.519 & 14.92 & 0.515 & 0.942 \\
$\alpha_2$ & 7 & 7.163 & 0.639 & 7.130 & 0.633 & 0.930 \\
$\sigma^2_{\epsilon}$ & 0.2 & 0.186 & 0.033 & 0.177 & 0.032 & 0.850\\
$\sigma^2_X$ & 0.2 & 0.229 & 0.050 & 0.235 & 0.052 & 0.896\\
$\rho$ & 0.9 & 0.853 & 0.049 & 0.862 & 0.041 & 0.850\\
{\it Logistic submodel} &&&&&&\\
$\beta_1$ & -22 & -19.28 & 3.578 & -19.561 & 3.043 & 0.996\\
$\beta_2$ & 5 & 5.15 & 0.935 & 5.218 & 0.797 & 0.998\\\hline
\end{tabular}
\end{table}

\section{Discussion}\label{disc}

In this paper we have proposed inferential strategies for a generalized linear model for a primary outcome with covariates that are underlying individual--specific random effects in a nonlinear random 
effects model for longitudinal data, considering correlated errors in the NLME. We use an MCMC procedure to jointly estimate all parameters in the model. The proposed approach provides a general framework for estimation in joint NLME--GLM models that circumvents any problem related with likelihood approximations.

In the analysis of the pregnancy dataset that motivates this work, we only use as the covariate for the logistic regression model the latent random effects of $\beta$--HCG profiles, but other covariates, such as 
age, number of previous normal and abnormal pregnancies and smoking status, could be useful for targeting specific 
individuals in future analysis. In our particular dataset, however, a number of women had 
missing values for many of these covariates. 

All the proposed estimators assume normality of random effects and 
within--individual errors. The latter is often reasonable, perhaps on a 
transformed scale. However, some authors \citep[e.g.,][among others]{VerbekeLesaffreJASA96}, 
 have shown that violation of this assumption can 
compromise inference in mixed--effects models, which raises similar concerns for the proposed joint 
model. Further research on methods that go beyond traditional 
normality assumption on random effects would be useful. These topics 
are the subject of current research to be reported elsewhere.

\section*{Acknowledgements}
We are grateful to Guillermo Marshall for facilitating us the $\beta$--HCG dataset.
Rolando de la Cruz thanks the Comisi\'on Nacional de 
Investigaci\'on Cient\'{\i}fica y Tecnol\'ogica -- CONICYT, Chile, for 
partially supporting his Ph.D. studies at the Pontificia Universidad 
Cat\'olica de Chile; Vicerrector\'{\i}a Adjunta de Investigaci\'on 
y Doctorado -- VRAID at the Pontificia Universidad Cat\'olica de Chile, 
for partially supporting this research under grant INICIO 06/2007;
Fondo Nacional de Desarrollo Cient\'{i}fico y Tecnol\'ogico
-- FONDECYT, Chile, for partially supporting this research under grant 1120739; and 
Programa de Investigaci\'on Asociativa -- PIA, CONICYT, 
for partially supporting this research under grant ANILLOS ACT--87.
Cristian Meza was supported by project FONDECYT 1141256 and grant ANILLOS ACT--1112, PIA, CONICYT, Chile. Ana Arribas-Gil was supported by projects MTM2010-17323 and ECO2011-25706, Spain. Carroll's research was supported by a grant from the
National Cancer Institute (R37-CA057030).

\bibliographystyle{plainnat}
\bibliography{refs}
\label{lastpage}
\end{document}